\documentclass[11pt,a4paper]{article}
\pdfoutput=1
\usepackage{jheppub}

\usepackage{color,epsfig} 
\usepackage{ifpdf}
\usepackage{amsmath} 
\usepackage{amssymb} 
\usepackage{bm}
\usepackage[english]{babel}
\usepackage{slashed}
\usepackage{multirow}

\usepackage{subfigure}

\title{Novel Leptoquark Pair Production at LHC}

\author[a,b]{Ilja Dor\v sner,} 
\author[c,d]{Svjetlana Fajfer,} 
\author[e]{and Ajla Lejli\' c} 

\affiliation[a]{University of Split, Faculty of Electrical Engineering, Mechanical Engineering and Naval Architecture in Split (FESB), R.\ Bo\v skovi\' ca 32, HR-21000 Split, Croatia}
\affiliation[b]{CERN, Theoretical Physics Department, CH-1211 Geneva 23, Switzerland}
\affiliation[c]{J.\ Stefan Institute, Jamova 39, P.\ O.\ Box 3000, SI-1001
  Ljubljana, Slovenia}
\affiliation[d]{Department of Physics,
  University of Ljubljana, Jadranska 19, SI-1000 Ljubljana, Slovenia}
\affiliation[e]{Department of Physics, Faculty of Science,
  University of Zagreb, Bijeni\v cka cesta 32, HR-10000 Zagreb, Croatia}
  
\emailAdd{dorsner@fesb.hr}
\emailAdd{svjetlana.fajfer@ijs.si} 
\emailAdd{ajla@me.com} 

\abstract{We introduce a novel mechanism for the leptoquark pair production at LHC that is of a $t$-channel topology and is quark-quark initiated. This mechanism operates under fairly general conditions. One of them is that the two leptoquarks in question couple to the same lepton and the other one is that the fermion numbers of these two leptoquarks  differ by two. The strength of the proposed mechanism provides an alternative way to the conventional processes to efficiently constrain the parameter space of the two leptoquark scenarios at LHC whenever the aforementioned conditions are met. We accordingly present one case study to outline the physics potential of this novel production mechanism.}

\arxivnumber{}

\begin{document}

\maketitle

\section{Introduction}
\label{sec:introduction}
Scalar leptoquarks exhibit very reach phenomenology and are frequently used either singly or in pairs to address some of the disagreements between the Standard Model predictions and experimental observations. For example, scalar leptoquarks can generate neutrino masses~\cite{Chua:1999si,Mahanta:1999xd,Dorsner:2017wwn,Cai:2017wry}, produce observable shifts in anomalous magnetic moments of charged leptons~\cite{Queiroz:2014zfa,Biggio:2014ela,ColuccioLeskow:2016dox,Crivellin:2018qmi,Kowalska:2018ulj,Mandal:2019gff,Dorsner:2019itg,Bigaran:2020jil,Dorsner:2020aaz}, provide a source of gauge coupling unification~\cite{Murayama:1991ah,Dorsner:2005fq,Dorsner:2005ii,Dorsner:2013tla,Cox:2016epl,Becirevic:2018afm}, and/or accommodate the $B$-physics anomalies~\cite{Kosnik:2012dj,Sakaki:2013bfa,Hiller:2014yaa,Sahoo:2015pzk,Pas:2015hca,Becirevic:2015asa,Bauer:2015knc,Cox:2016epl,Das:2016vkr,Popov:2016fzr,Becirevic:2016yqi,Dorsner:2017ufx,Chen:2017hir,Buttazzo:2017ixm,Angelescu:2018tyl,Azatov:2018kzb,Marzocca:2018wcf,Kumar:2018kmr,Becirevic:2018afm,Crivellin:2019dwb,Popov:2019tyc,Saad:2020ihm,Saad:2020ucl}. There is thus a strong physics case to search for the scalar leptoquark states in order to probe the relevant parameter space of these hypothetical particles.

\begin{figure}[b]
\centering
\includegraphics[scale=0.73]{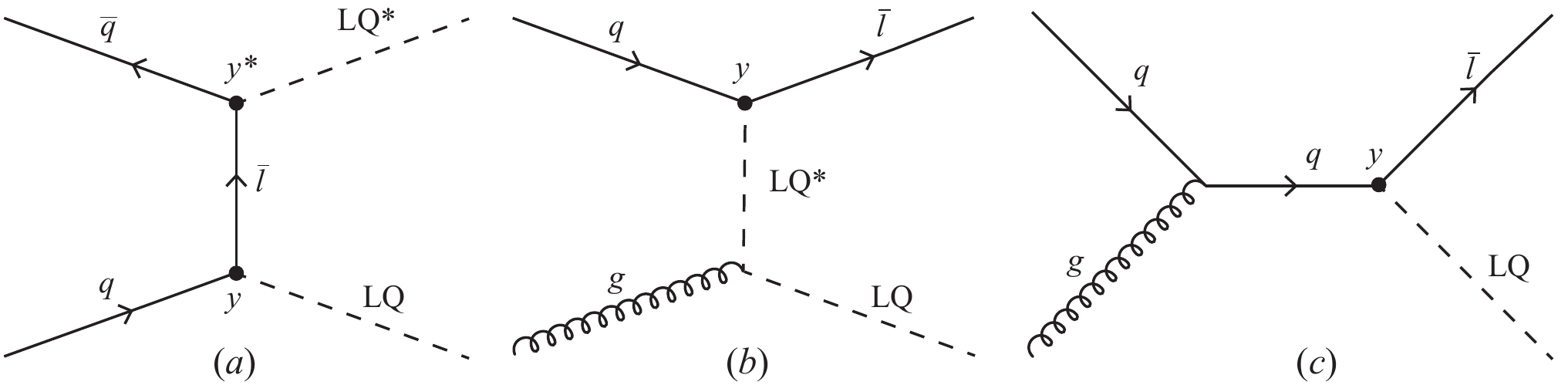}
\caption{\label{fig:DIAGRAM_a} Panel $(a)$ shows a $t$-channel Feynman diagram relevant for a pair production of leptoquarks at LHC. Panels $(b)$ and $(c)$ feature contributions towards a single leptoquark production. Here, $y$ represents Yukawa coupling of a lepton ($l$) and a quark ($q$) with a leptoquark (LQ).}
\end{figure}

The CMS and ATLAS analyses of the leptoquark pair production signatures at LHC are currently some of the most robust sources of experimental constraints on the leptoquark parameter space. (See, for example, Refs.~\cite{CMS:2020gru,Aad:2020iuy,Aad:2020jmj,Sirunyan:2020zbk,Aad:2021rrh} for a sample of the latest bounds.) The relevant cross sections for the leptoquark pair production in proton-proton collisions are already available at the next-to-leading order~\cite{Kramer:1997hh,Kramer:2004df,Mandal:2015lca,Dorsner:2018ynv,Borschensky:2020hot} as well as the next-to-next-to-leading order~\cite{Beenakker:2016lwe,Beenakker:1997ut,Beenakker:2010nq,Beenakker:2016gmf} in strong coupling constant and the associated tools to produce these cross sections are readily accessible for the signal simulation if and when needed. Apart from the QCD driven contribution towards the pair production cross section at hadron colliders there is only one diagram that exhibits any dependence on the leptoquark Yukawa coupling(s) to the quark-lepton pairs. It is of the $t$-channel type, as shown in panel $(a)$ of Fig.~\ref{fig:DIAGRAM_a}, where the initial state is always made out of a quark-antiquark combination. 

As it turns out, the $t$-channel contribution of Fig.~\ref{fig:DIAGRAM_a} towards the leptoquark pair production at LHC becomes relevant only when the leptoquark interacts very strongly, i.e., with the couplings of order one, to the quarks of the first generation~\cite{Dorsner:2014axa}. However, in this particular regime other processes such as a single leptoquark production~\cite{Dorsner:2014axa,Alves:2002tj,Hammett:2015sea,Mandal:2015vfa}, induced by the diagrams in panels $(b)$ and $(c)$ of Fig.~\ref{fig:DIAGRAM_a}, the Drell-Yan di-lepton production~\cite{Faroughy:2016osc,Raj:2016aky,Greljo:2017vvb,Bansal:2018eha,Schmaltz:2018nls,Fuentes-Martin:2020lea}, and/or a resonant leptoquark production~\cite{Ohnemus:1994xf,Eboli:1997fb,Buonocore:2020erb,Greljo:2020tgv} might become equally capable of or even more important in constraining the leptoquark parameter space. Moreover, one also needs to take into account the experimental constraints from flavor physics, such as the Atomic Parity Violation ones, for a given $l$-$q$-LQ scenario to ensure its viability, especially within a large Yukawa coupling regime~\cite{Dorsner:2016wpm}. (For an example of a combined analysis of the relevant experimental constraints with regard to the potential viability of the leptoquark parameter space see Ref.~\cite{Crivellin:2021egp}.) It is thus often stated that the pair production of leptoquarks at LHC, within the otherwise viable parameter space, is purely QCD driven. This is assumed to hold even when a given leptoquark multiplet has more than one electric charge component~\cite{Diaz:2017lit} and/or when there is more than one leptoquark multiplet present within a given New Physics scenario~\cite{Becirevic:2018afm}. 
It is our intention to show that one needs to be careful with this assumption, especially in the latter case. We will demonstrate this within one explicit flavor construction just to prove our point. It is straightforward to adopt our analysis to address other scenarios with the same underlying features that will be clearly outlined in our study. 

The main premise behind our work is a possibility that one might have a New Physics scenario that allows for a production of two scalar leptoquarks via the $t$-channel type of contribution of Fig.~\ref{fig:DIAGRAM_a} but with two quarks instead of the quark-antiquark pair in the initial state, where the two quarks do not have to necessarily be of the same flavor. The main benefits of that possibility, for the leptoquark pair production at LHC, are $(i)$ the potential absence of the parton distribution function (PDF) suppression of the associated cross section and $(ii)$ the introduction of additional production channels. The quark-quark initiated contribution can thus start to dominate over the QCD production for much smaller values of Yukawa couplings when compared to the quark-antiquark initiated $t$-channel case. Moreover, it can even start to dominate over the single leptoquark production due to a more pronounced scaling, i.e., quartic vs.\ quadratic, with respect to the Yukawa couplings. The quark-quark initiated $t$-channel contribution thus has a potential to be even more important than the single leptoquark production, the Drell-Yan di-lepton production, and/or the resonant leptoquark production in constraining certain parts of the parameter space spanned by the relevant Yukawa couplings and leptoquark masses.

\begin{figure}[b]
\centering
\includegraphics[scale=0.75]{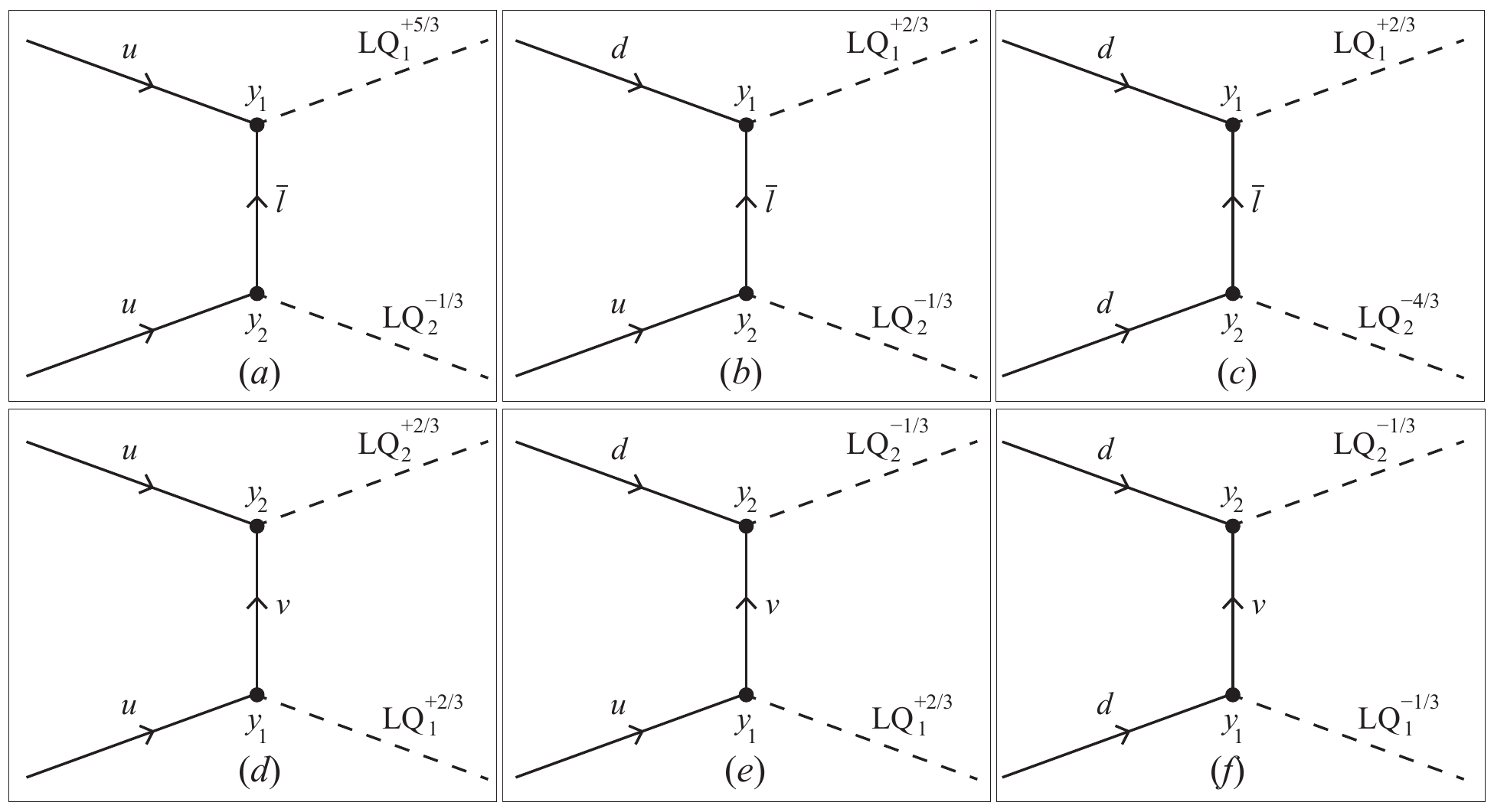}
\caption{\label{fig:DIAGRAM_b} The leading order Feynman diagrams for the novel leptoquark pair production at LHC, where the final state leptoquarks $\mathrm{LQ}_1$ and $\mathrm{LQ}_2$ have fermion numbers that differ by two. The electric charges of the two leptoquarks, in units of an absolute value of the electron charge, are denoted with superscripts.}
\end{figure}

Before we present our case study we outline the general features of the processes we aim to investigate. The relevant diagrams are shown in Fig.~\ref{fig:DIAGRAM_b}. The $u$ and $d$ in Fig.~\ref{fig:DIAGRAM_b} are the up quark and the down quark, respectively, and we explicitly denote the electric charges of the two final state leptoquarks with superscripts. We will concentrate our attention, in Sec.~\ref{sec:case_study}, on one particular scenario that induces both the $uu$ and $ud$ diagrams as shown in panels $(a)$ and $(b)$ of Fig.~\ref{fig:DIAGRAM_b}, respectively. Of course, one might also need to look at the contributions from other initial state quark flavor combinations within a given leptoquark scenario.

It is clear from Fig.~\ref{fig:DIAGRAM_b} that, due to a conservation of the fermion number $F=3 B+L$, where $B$ and $L$ are the baryon and lepton numbers, respectively, the final state should comprise one leptoquark with $F=0$ and the other one with $|F|=2$. Concretely, $\mathrm{LQ}_1$ and $\mathrm{LQ}_2$ of Fig.~\ref{fig:DIAGRAM_b} are $F=0$ and $|F|=2$ scalar fields, respectively. In the leptoquark parlance, the production processes we intend to analyse in this work can occur whenever the New Physics scenario combines $S_3$, $\tilde{S}_1$ or $S_1$ with either $R_2$ or $\tilde{R}_2$. We stress that the leptoquark pairs in question do not need to mix via the Higgs boson for this production mechanism to work. 

With these preliminaries out of the way, we turn our attention to the case study.

\section{Case Study}
\label{sec:case_study}

We study in detail one particular flavor realisation of the $S_1$-$R_2$ leptoquark pair scenario. The relevant interaction terms in the Lagrangian are
\begin{equation}
\label{eq:main_S_1_R_2}
\mathcal{L} \supset +y_{1\,ij}\bar{e}_{R}^{i} R_{2}^{a\,*}Q_{L}^{j,a}+y_{2\,ij}\bar{u}_{R}^{C\,i} S_{1} e_{R}^{j} +\textrm{h.c.},
\end{equation}
where $y_1$ and $y_2$ are Yukawa coupling matrices in the flavor space and $a$ is the $SU(2)$ index. We take for simplicity that both electric charge eigenstates in $R_2$ are mass degenerate. We furthermore assume that the fields comprising $R_2$ have the same mass as $S_1$ leptoquark, if simultaneously present, and we denote this common mass parameter with $m_\mathrm{LQ}$ in our study. 

If we expand Eq.~\eqref{eq:main_S_1_R_2} in terms of the leptoquark and fermion mass eigenstates we obtain 
\begin{equation}
\label{eq:main}
\mathcal{L} \supset + (y_1 V^\dagger)_{ij} \bar{e}^i_R 
  u^j_L R_2^{+5/3\,*} +y_{1\,ij} \bar{e}^i_R d^j_L R_2^{+2/3\,*} +y_{2\,ij}\bar{u}_{R}^{C\,i} S^{+1/3}_{1} e_{R}^{j}+ \textrm{h.c.},
\end{equation}
where $V$ represents a Cabibbo--Kobayashi--Maskawa (CKM) mixing matrix. The leptoquark superscript in Eq.~\eqref{eq:main} denotes its electric charge in units of an absolute value of the electron charge. To simplify things even further we assume that the only entry in $y_1$ matrix that is different from zero is the 21 element while in $y_2$ we take the only non-zero entry to be the 12 element. If both of these couplings are present, we take them to be equal to each other and denote that parameter with $y$, i.e., $y_{1\,21}=y_{2\,12}=y$, where $y$ is always assumed to be real. Finally, we set the CKM matrix to be equal to the identity matrix in our numerical simulations. The departure from the CKM assumption will only make a tiny quantitative difference without any change of the qualitative picture we are interested in. With all these assumptions we have a flavor scenario where all three leptoquarks, i.e., $S^{+1/3}_{1}$, $R_2^{+5/3}$, and $R_2^{+2/3}$, couple exclusively to muons. The only difference is that $S^{+1/3}_{1}$ and $R_2^{+5/3\,*}$ couple to an up quark whereas $R_2^{+2/3\,*}$ couples to a down quark. Since $R_2$ and $S_{1}$ are $F=0$ and $|F|=2$ leptoquarks, respectively, and since they both couple to the same lepton, we have all the necessary ingredients to generate the pair production processes we are interested in. In particular, this scenario has an ability to simultaneously generate the diagrams given in panels $(a)$ and $(b)$ of Fig.~\ref{fig:DIAGRAM_b}, where both leptoquarks in the final state, i.e., $\mathrm{LQ}_1$ and $\mathrm{LQ}_2$, will exclusively decay into muons and jets, where the jets are initiated by either $u$ or $d$ quarks and are hence indistinguishable from the experimental point of view. Again, the final state for the leptoquark pair production, within our flavor ansatz, will always comprise a muon-antimuon pair and two jets.  

We are finally ready to make some quantitative comparisons. In order to do that we implement the aforementioned $S_1$-$R_2$ scenario in the universal {\rmfamily\scshape FeynRules} output format~\cite{Alloul:2013bka} and 
generate relevant cross sections within the {\rmfamily\scshape MadGraph5\_aMC@NLO} framework~\cite{Alwall:2014hca} using the {\tt nn23lo1} PDF set~\cite{Ball:2014uwa}. All our simulations are of the leading order nature to allow for fair and simple comparisons. The center-of-mass energy for proton-proton collisions is set at 13\,TeV  and we always use central values for relevant production cross sections using fixed factorisation ($\mu_F$) and renormalisation scales ($\mu_R$) set at $\mu_F=\mu_R= m_\mathrm{LQ}/2$.

Our analysis follows closely the approach advocated in Ref.~\cite{Dorsner:2014axa}. Namely, we use the fact that the functional dependence for the conventional leptoquark pair production at LHC, within our flavor construction, can be written as
\begin{equation}
\label{eq:pair_c_s}
\sigma_{\textrm{P}}(y,m_\mathrm{LQ})= a_0(m_\mathrm{LQ})+a_2(m_\mathrm{LQ}) y^2+a_4 (m_\mathrm{LQ}) y^4.
\end{equation} 
The first term in Eq.~\eqref{eq:pair_c_s} represents the QCD pair production contribution towards cross section. It only depends on the mass of the leptoquark for a given center-of-mass energy and the associated PDF set. The third term in Eq.~\eqref{eq:pair_c_s} is the $t$-channel production contribution whereas the second term is the interference between the $t$-channel and the QCD amplitudes. It is important to note that both $a_2(m_\mathrm{LQ})$ and $a_4(m_\mathrm{LQ})$ depend on the flavor of the quark that the leptoquark couples to. Clearly, it is sufficient to numerically evaluate the pair production cross section for three different values of Yukawa coupling $y$ for a fixed value of $m_\mathrm{LQ}$ in order to extract information on $a_0(m_\mathrm{LQ})$, $a_2(m_\mathrm{LQ})$, and $a_4(m_\mathrm{LQ})$ and thus provide an input for subsequent interpolation. To accommodate the latest bounds on the leptoquark mass we start this process for $m_\mathrm{LQ}=1.6$\,TeV and proceed with our numerical simulation up to $m_\mathrm{LQ}=2.4$\,TeV in the 200\,GeV increments. (A leptoquark that couples exclusively to the first generation quarks and a muon cannot have a mass below 1.7\,TeV~\cite{Aad:2020iuy}.) Note that $S^{+1/3}_{1}$ and $R_2^{+5/3}$ will have exactly the same functional dependence for $\sigma_{\textrm{P}}(y,m_\mathrm{LQ})$ since they both couple purely to $u$ quark. We accordingly denote this cross section with $\sigma_{\textrm{P}u}$. The cross section $\sigma_{\textrm{P}}(y,m_\mathrm{LQ})$ to produce a $R_2^{+2/3}$-$R_2^{+2/3\,*}$ pair is analogously denoted with $\sigma_{\textrm{P}d}$. Note, again, that the QCD contribution, given by $a_0(m_\mathrm{LQ})$ of Eq.~\eqref{eq:pair_c_s}, will be the same regardless whether one looks at $pp \rightarrow S_1^{+1/3}S^{+1/3\,*}_1$, $pp \rightarrow R_2^{+5/3}R_2^{+5/3\,*}$, or $pp \rightarrow R_2^{+2/3}R_2^{+2/3\,*}$ processes.

The cross section for the single leptoquark production, on the other hand, scales quadratically with respect to the Yukawa coupling as 
\begin{equation}
\label{eq:single_c_s}
\sigma_{\textrm{S}}(y,m_\mathrm{LQ})= a(m_\mathrm{LQ}) y^2,
\end{equation} 
where $a(m_\mathrm{LQ})$ encapsulates its mass and flavor dependence. We accordingly denote with $\sigma_{\textrm{S}u}$ the single leptoquark production cross section for $pp \rightarrow S^{+1/3}_{1} \mu^-$ together with $pp \rightarrow S^{+1/3\,*}_{1} \mu^+$. The same cross section is applicable for $pp \rightarrow R_2^{+5/3} \mu^-$ together with $pp \rightarrow R_2^{+5/3\,*} \mu^+$. For the single leptoquark production of $R_2^{+2/3}$, on the other hand, we introduce $\sigma_{\textrm{S}d}$. To numerically extract information on the leptoquark mass and flavor dependence of $a(m_\mathrm{LQ})$ we evaluate $\sigma_{\textrm{S}}(y,m_\mathrm{LQ})$ for aforementioned leptoquark mass range at the fixed value of $y$ and perform an interpolation.

Finally, when we consider the $uu$ contribution of panel $(a)$ from Fig.~\ref{fig:DIAGRAM_b} towards the pair production cross section that corresponds to the $t$-channel production of a $R_2^{+5/3}$-$S^{+1/3\,*}_{1}$ pair we denote it with $\sigma_{uu}$ whereas the cross section for the production of a $R_2^{+2/3}$-$S^{+1/3\,*}_{1}$ pair that corresponds to the process of panel $(b)$ from Fig.~\ref{fig:DIAGRAM_b} is denoted with $\sigma_{ud}$. Both the $uu$ and $ud$ contributions individually scale as
\begin{equation}
\sigma_{uu\,(ud)}= a_{uu\,(ud)}(m_\mathrm{LQ}) y^4,
\end{equation}
where the functional dependence of $a_{uu}(m_\mathrm{LQ})$ and $a_{ud}(m_\mathrm{LQ})$ is obtained by evaluating the cross sections $\sigma_{uu}$ and $\sigma_{ud}$ for $pp \rightarrow R_2^{+5/3}S^{+1/3\,*}_{1}$ and $pp \rightarrow R_2^{+2/3}S^{+1/3\,*}_{1}$, respectively, for aforementioned leptoquark mass range at the fixed value of $y$. 

\begin{figure}[b]
\centering
\includegraphics[scale=0.73]{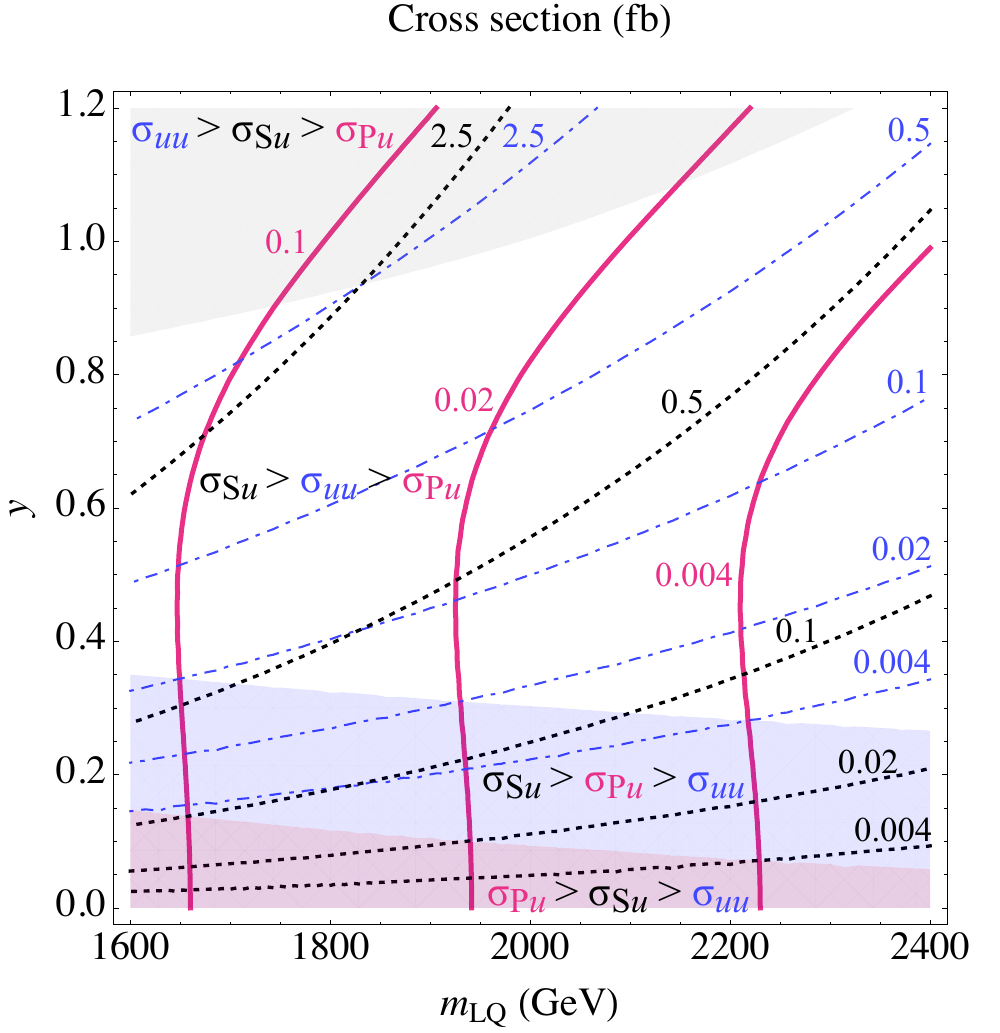}
\includegraphics[scale=0.73]{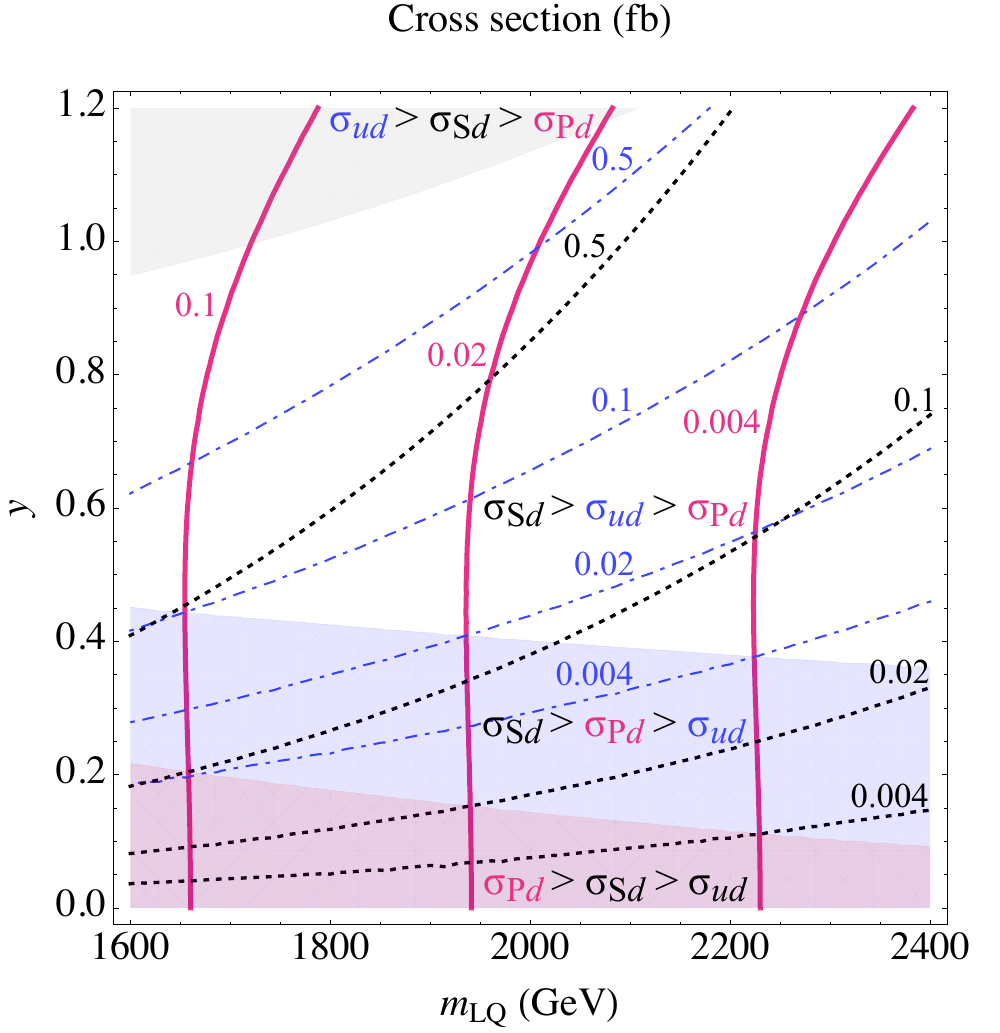}
\caption{\label{fig:DIAGRAM_c} The contours of constant value for the novel contributions ($\sigma_{uu\,(ud)}$) and the usual pair ($\sigma_{\textrm{P}u\,(\textrm{P}d)}$) and single leptoquark production ($\sigma_{\textrm{S}u\,(\textrm{S}d)}$) cross sections as a function of $m_\mathrm{LQ}$ and coupling $y$. Colored regions correspond to the particular orderings of the cross sectional strengths.}
\end{figure}

Our initial intent is just to show the strength of $\sigma_{uu\,(ud)}$ contributions when compared to the usual pair and single leptoquark production cross sections. To that end we show in Fig.~\ref{fig:DIAGRAM_c} the contours of constant value for the aforementioned cross sections as a function of $m_\mathrm{LQ}$ and Yukawa coupling(s) $y$. 
The left panel of Fig.~\ref{fig:DIAGRAM_c} features the production mechanisms that are associated with the states that couple to the up quarks, i.e., $S^{+1/3}_{1}$ and $R_2^{+5/3}$, while the right panel concerns the production mechanisms for the state that couples to the down quarks, i.e., $R_2^{+2/3}$. The only intricacy here is that in order to have the $\sigma_{ud}$ contribution one needs the presence of both $S^{+1/3}_{1}$ and $R_2^{+2/3}$, as shown in panel $(b)$ of Fig~\ref{fig:DIAGRAM_b}. The solid lines in Fig.~\ref{fig:DIAGRAM_c} represent the contours of constant value for cross section of the conventional pair production, the dashed contours correspond to the constant values of the single leptoquark production cross sections whereas the dot-dashed contours correspond to the constant cross sections for the novel pair production via quark-quark initial state. We also specify the values of the actual cross sections in femtobarn units in Fig.~\ref{fig:DIAGRAM_c}. 

It is clear from the left panel of Fig.~\ref{fig:DIAGRAM_c} that, within the leptoquark mass range we consider, there are four distinct regions that correspond to the following situations, going from the bottom to the top of that panel: $\sigma_{\textrm{P}u}>\sigma_{\textrm{S}u}>\sigma_{uu}$, $\sigma_{\textrm{S}u}>\sigma_{\textrm{P}u}>\sigma_{uu}$, $\sigma_{\textrm{S}u}>\sigma_{uu}>\sigma_{\textrm{P}u}$, and $\sigma_{uu}>\sigma_{\textrm{S}u}>\sigma_{\textrm{P}u}$. It is the single leptoquark production that starts to dominate first over the pair production. But, it is soon followed by the $\sigma_{uu}$ contribution. In fact, as advocated at the beginning of this manuscript, our simulation shows that there is even a region where the novel contribution encapsulated solely by $\sigma_{uu}$ starts to dominate over all the conventional production mechanisms. The exact same qualitative picture holds for the scenario that concerns the leptoquarks that couple to the down quarks as can be seen in the right panel of Fig.~\ref{fig:DIAGRAM_c}. What is important to notice is that $\sigma_{uu}$ and $\sigma_{ud}$ represent individual contributions towards the pair production cross section which simply means that the overall enhancement effect can and, in our case study, will be much more pronounced, as we show next.

\begin{figure}
\centering
\includegraphics[scale=0.73]{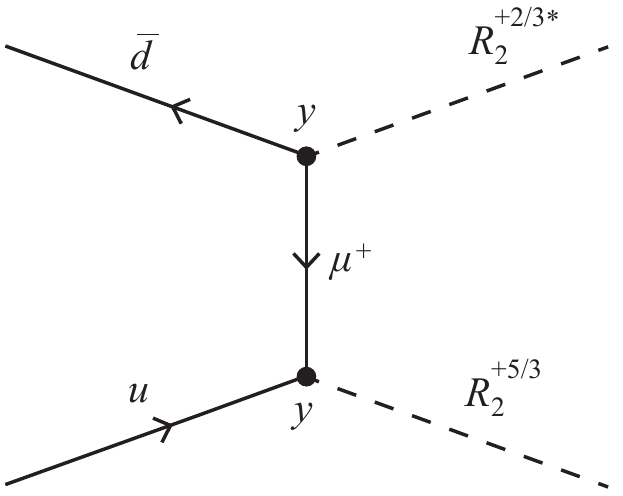}
\caption{\label{fig:DIAGRAM_d} A self-induced $t$-channel contribution towards a pair production of the $R_2$ leptoquark multiplet components at LHC within our flavor ansatz.}
\end{figure}

What remains to be seen is how the pair production cross sections compare for different scenarios. Namely, we want to see what the difference is between the case when only $S_1$ leptoquark is present, when only $R_2$ is present, and, finally, when both $S_1$ and $R_2$ are simultaneously present. Note that in the $R_2$ case one needs to take into consideration an additional process that is depicted in Fig.~\ref{fig:DIAGRAM_d} for our particular flavor ansatz. There are thus four contributions towards the leptoquark pair production in the $R_2$ scenario, i.e., $pp \rightarrow R_2^{+5/3}R_2^{+5/3\,*}$, $pp \rightarrow R_2^{+2/3}R_2^{+2/3\,*}$, $pp \rightarrow R_2^{+5/3}R_2^{+2/3\,*}$, and $pp \rightarrow R_2^{+5/3\,*}R_2^{+2/3}$, that one needs to consider.
In the $S_1$-$R_2$ scenario one has to account for the additional production modes such as $R_2^{+5/3}$-$S^{+1/3\,*}_{1}$ and $R_2^{+2/3}$-$S^{+1/3\,*}_{1}$ as well as the associated charge conjugate terms. Note, once again, that upon decay one would, in all three scenarios, have two opposite sign muons and two jets in the final state. So, this comparison will yield an accurate estimate of the overall effect of the proposed mechanism within this flavor ansatz.

Before we present the results of our numerical analysis, let us briefly discuss the naive expectation for these pair production cross sections that we denote with $\sigma_{\textrm{P}\,S_1}$, $\sigma_{\textrm{P}\,{R_2}}$, and $\sigma_{\textrm{P}\,{S_1\textnormal{-}R_2}}$. To that end we introduce a dimensionless parameter $\mu_\mathrm{LQ}$ that is defined as the ratio of an actual pair production cross section for a given leptoquark scenario and the QCD driven pair production for a single leptoquark scenario, as encapsulated by $a_0(m_\mathrm{LQ})$ of Eq~\eqref{eq:pair_c_s}, for a fixed leptoquark mass. If the cross section does not depend on Yukawa coupling, or Yukawa coupling is small, one would expect to get $\mu_{S_1}=1$. Of course, following the same reasoning, one would expect that $\mu_{R_2}=2$ since $R_2$ is made out of two physical mass-degenerate leptoquark states. Finally, one would naively expect to find $\mu_{S_1\textnormal{-}R_2}=3$ for the $S_1$-$R_2$ scenario if $S_1$ and the two states in $R_2$  are degenerate in mass. 

\begin{figure}[b]
\centering
\includegraphics[scale=0.73]{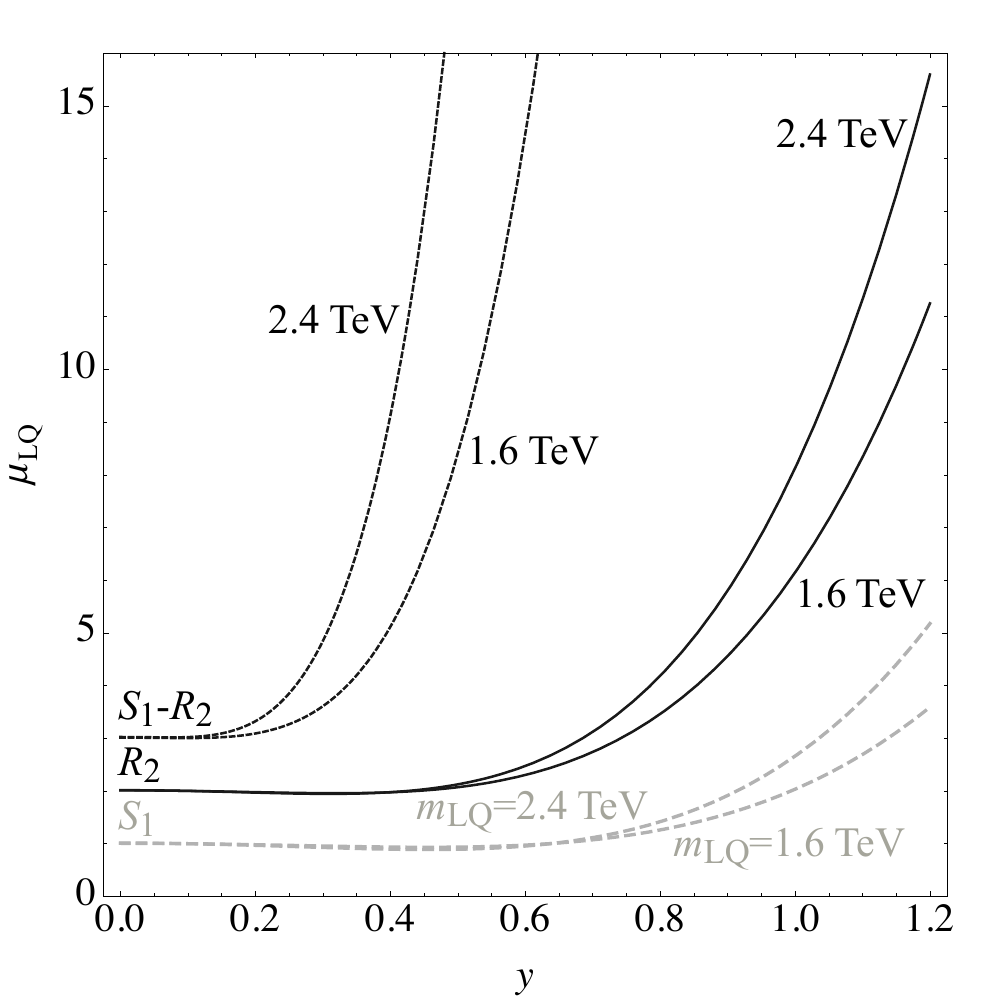}
\includegraphics[scale=0.73]{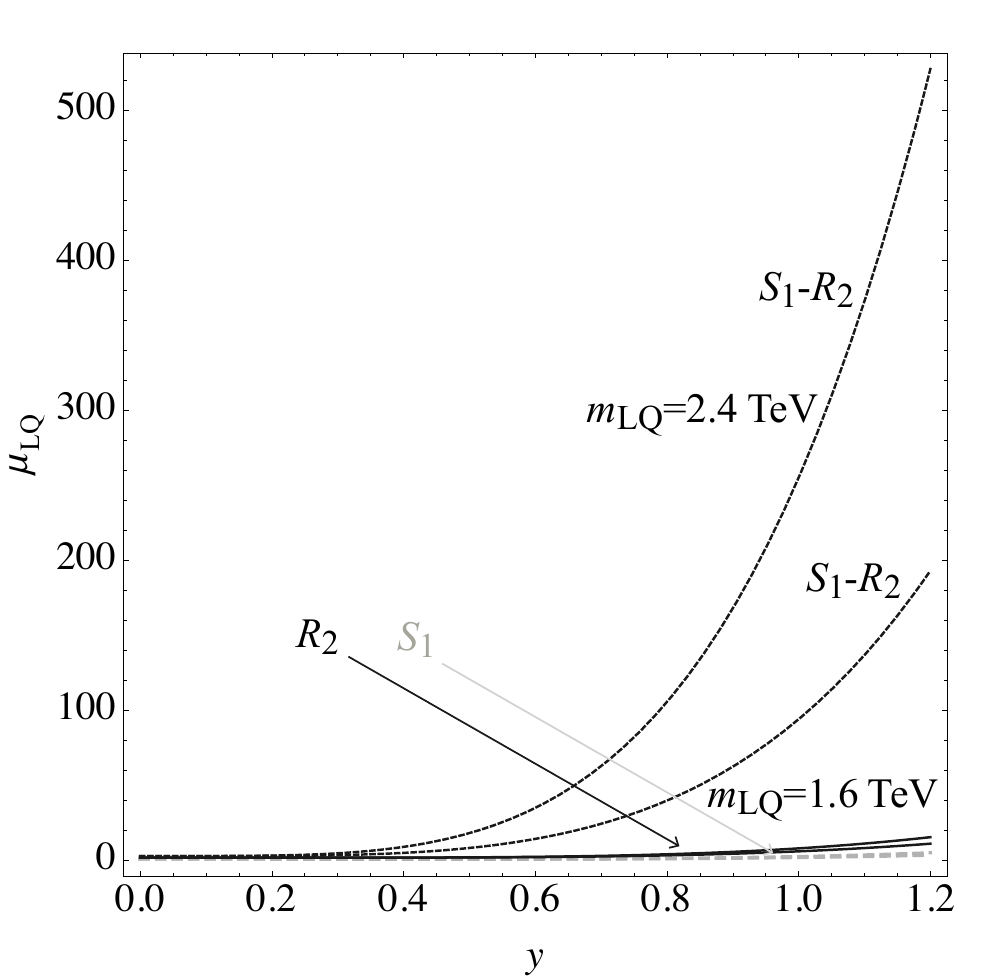}
\caption{\label{fig:DIAGRAM_e} Ratios of the factual cross sections for the pair production of leptoquarks at LHC for three scenarios --- $S_1$, $R_2$, and $S_1$-$R_2$ --- and the QCD cross section for the pair production of one leptoquark as a function of the associated Yukawa coupling(s). In all three instances the leptoquarks of mass $m_\mathrm{LQ}$, as indicated in the plots, couple to the first generation quarks and a right-chiral muon with Yukawa coupling strength $y$ while the center-of-mass energy of the $pp$ collisions is 13\,TeV.}
\end{figure}

The outcome of our numerical analysis for $\mu_{S_1}$, $\mu_{R_2}$, and $\mu_{S_1\textnormal{-}R_2}$ is shown in Fig.~\ref{fig:DIAGRAM_e}. Indeed, for small $y$ one does observe the naive scaling of $\mu_\mathrm{LQ}$. However, this behaviour breaks down for relatively small values of Yukawa couplings, as can be seen from the left panel of Fig.~\ref{fig:DIAGRAM_e}. In fact, the breaking of the naive picture is especially strong in the case of the $S_1$-$R_2$ scenario, where the novel contributions towards pair production that correspond to the Feynman diagrams shown in panels $(a)$ and $(b)$ of Fig.~\ref{fig:DIAGRAM_b} start to kick in at about $y \approx 0.1$. Since the novel contributions scale with the fourth power of Yukawa coupling, the effect is very strong at large $y$ and we accordingly present its full range in the right panel of Fig.~\ref{fig:DIAGRAM_e}. The enhancement of $\mu_\mathrm{LQ}$ parameter is more pronounced for larger values of $m_\mathrm{LQ}$ since the phase space suppression for the conventional pair production is more relevant in that regime. We accordingly show in Fig.~\ref{fig:DIAGRAM_e} behaviour of $\mu_{S_1}$, $\mu_{R_2}$, and $\mu_{S_1\textnormal{-}R_2}$ for two different leptoquark masses, i.e., $m_\mathrm{LQ}=1.6$\,TeV and $m_\mathrm{LQ}=2.4$\,TeV. For example, in the $S_1$ case, at $y=1.0$, one would see a signal that is approximately 2 times larger than the naive expectation. For the $R_2$ case, at $y=1.0$, a signal would exceed naive expectation by a factor of 4 for $m_\mathrm{LQ}=2.4$\,TeV. However, in the $S_1$-$R_2$ case, at $y=1.0$, one would see a signal that is approximately 30 times larger than the expected value for $m_\mathrm{LQ}=1.6$\,TeV and approximately 80 times larger for $m_\mathrm{LQ}=2.4$\,TeV. And, for $y=1.2$, all the aforementioned enhancements would double with respect to the $y=1.0$ case due to the quartic nature of the effect. As advocated, one needs to be rather careful if more than one leptoquark multiplet is present to make sure that all the relevant contributions towards the scalar leptoquark pair production are taken into account. 
We specify, for the benefit of the reader, the cross sections associated with Fig.~\ref{fig:DIAGRAM_e} in Table~\ref{table:c_s}.  
\begin{table}
\begin{center}
\begin{tabular}{| c | c | c | c | c |}\hline
$m_\textrm{LQ}$ (TeV) & $y$ & $\sigma_{\textrm{P}\,S_1}$ (fb) & $\sigma_{\textrm{P}\,R_2}$ (fb) & $\sigma_{\textrm{P}\,S_1\textnormal{-}R_2}$ (fb)\\\hline\hline
& $0.1$ & $0.141$ & $0.283$ & $0.425$\\\cline{2-5}
1.6 & $0.5$ & $0.132$ & $0.293$ & $1.19$\\\cline{2-5}
& $1.0$ & $0.287$ & $0.869$ & $13.3$\\\hline
\hline
& $0.1$ & $0.0143$ & $0.0286$ & $0.0432$\\\cline{2-5}
2.0 & $0.5$ & $0.0131$ & $0.0299$ & $0.177$\\\cline{2-5}
& $1.0$ & $0.0330$ & $0.100$ & $2.27$\\\hline
\hline
& $0.1$ & $0.00155$ & $0.00311$ & $0.00469$\\\cline{2-5}
2.4 & $0.5$ & $0.00139$ & $0.00330$ & $0.0284$\\\cline{2-5}
& $1.0$ & $0.00413$ & $0.0126$ & $0.396$\\
\hline\end{tabular}
\end{center}
\caption{The leading order cross sections for the leptoquark pair production within the $S_1$, $R_2$, and $S_1$-$R_2$ scenarios through the proton-proton collisions at 13\,TeV center-of-mass energy, where in all three instances leptoquarks of mass $m_\textrm{LQ}$ couple exclusively to a right-chiral muon and the first generation quarks, as allowed by the Standard Model gauge group, with the coupling strength $y$.}
\label{table:c_s}
\end{table}

Before we conclude some final remarks are in order. The main enhancement of the pair production in our case study originates from the diagrams in panels $(a)$ and $(b)$ of Fig.~\ref{fig:DIAGRAM_b}. However, there could be additional contributions from other diagrams in Fig.~\ref{fig:DIAGRAM_b} within a given two leptoquark scenario. The point is that the effect we advocate could be even larger than what we have presented in Fig.~\ref{fig:DIAGRAM_e}. Also, it could be possible to construct a flavor scenario where the enhancement is through, for example, $uu$ mode but the two final state leptoquarks decay preferentially into quarks of heavier flavor(s) and/or into different lepton(s) than the one that is exchanged in the $t$-channel. Moreover, the two final state leptoquarks do not have to be mass degenerate which is often one of the underlying assumptions to reconstruct the source of the final state particles in the data analysis of the pair production at LHC. This would open up other kinematically distinct signatures for the final state leptons and quarks. This mechanism can lead to enhancement even when the initial state is not associated with the valence quarks simply because it allows for additional production channels. Again, there are multiple effects of different nature that are associated with the proposed pair production mechanism that certainly deserve additional attention within a well-defined leptoquark scenario. Finally, we have noticed that the additional self-induced contribution towards the leptoquark pair production, as given by the Feynman diagram of Fig.~\ref{fig:DIAGRAM_d} for the $R_2$ scenario, has not been considered in the literature at all. We accordingly plan to purse the associated phenomenology in future publications~\cite{DFLS}.

\section{Conclusions}
\label{sec:conclusions}

We introduce and discuss a novel production mechanism to generate scalar leptoquarks in pairs at LHC. The proposed mechanism is based on the $t$-channel Feynman diagram topology and it has quark-quark pairs in the initial state which makes it perfectly suited for the LHC studies. The mechanism in question requires a presence of two scalar leptoquarks that couple to the same lepton and whose fermion numbers differ by two. So, the mechanism can be operational whenever the New Physics scenario combines $S_3$, $\tilde{S}_1$ or $S_1$ with either $R_2$ or $\tilde{R}_2$. 

To demonstrate the strength of this mechanism we study one particular flavor realisation based on the $S_1$-$R_2$ scenario, where both $S_1$ and $R_2$ couple to a right-chiral muon and the quarks of the first generation, as governed by the Standard Model gauge group. We explicitly show that the pair production signal at LHC can be easily enhanced by a factor of ten to a hundred even for the moderate values of the Yukawa couplings when compared with the naive expectation as given by the QCD driven production mechanisms. It can thus serve as an alternative way to the conventional processes to efficiently constrain the parameter space of the two leptoquark scenarios at LHC within otherwise allowed parameter space spanned by the Yukawa couplings and leptoquark masses.

\acknowledgments

I.D.\ would like to thank the CERN Theory Department and Institute Jo\v{z}ef Stefan on hospitality. The work of S.F.\ was in part financially supported by the Slovenian Research Agency (research core funding No.\ P1-0035).

\end{document}